\documentclass[reprint, amsmath,amssymb,aps,longbibliography]{revtex4-2}

\usepackage[bottom]{footmisc}% places footnotes at page bottom

\usepackage{kpfonts}
\usepackage[T1]{fontenc}
\usepackage[latin9]{inputenc}
\usepackage{mathrsfs}
\usepackage{bm}
\usepackage{amsmath}
\usepackage{amsthm}
\usepackage{amssymb}
\usepackage{stmaryrd}
\usepackage{mathtools}
\usepackage{changepage}
\usepackage{lineno}
\usepackage{comment}
\usepackage{graphicx}% Include figure files
\usepackage{mathtools}
\usepackage{dcolumn}% Align table columns on decimal point
\usepackage{bm}% bold math
\usepackage{xr}
\DeclarePairedDelimiter\abs{\lvert}{\rvert}%
\makeatletter
\let\oldabs\abs
\def\abs{\@ifstar{\oldabs}{\oldabs*}}

\DeclareFontShape{OMX}{cmex}{m}{b}{<-> cmexb10}{}
\SetSymbolFont{largesymbols}{bold}{OMX}{cmex}{m}{b}

\DeclarePairedDelimiterX{\KLDx}[2]{(}{)}{%
  #1 \parallel #2%
}
\newcommand{\KLD}{D\KLDx}

\usepackage{stackengine}
\stackMath

\edef\ordinarycolon{\mathchar\the\mathcode`: }
\edef\ordinaryequals{\mathchar\the\mathcode`= }

\usepackage{breqn}
\AtBeginDocument{%
	\catcode`^=12
}

\allowdisplaybreaks

\let\cat@comma@active\@empty

\usepackage{color}

\usepackage{hyperref}
\usepackage[capitalize]{cleveref}

\newif\ifnotes
\notestrue

\usepackage{color}% for colors
\RequirePackage[dvipsnames,usenames]{xcolor}

\makeatother

\newcommand{\ba}{\begin{eqnarray}}
	\newcommand{\ea}{\end{eqnarray}}

\newcommand{\eq}[1]{\begin{align}#1\end{align}}
\newcommand{\spl}[1]{\begin{split}#1\end{split}}

\newcommand{\lr}[1]{\langle #1 \rangle}

\newcommand{\V}{{\mathcal{V}}}

\usepackage{color}
\definecolor{myblue}{rgb}{.8, .8, 1}

\usepackage{amsmath}
\usepackage{empheq}

\newlength\mytemplen
\newsavebox\mytempbox

\begin{document}
	
\preprint{APS/123-QED}

%\title{The fundamental thermodynamic costs of communication}% 

\title{Entropy production in communication channels}

\author{Farita Tasnim}
\email[]{farita@mit.edu, web: farita.me}% Your name
\affiliation{Massachusetts Institute of Technology, Cambridge, MA, USA}

    \author{Nahuel Freitas}
\email[]{nahuel.freitas@uni.lu}% Your name
\affiliation{Departamento de Fisica, FCEyN, UBA, Pabellon 1, Ciudad Universitaria, 1428 Buenos Aires, Argentina}

\author{David H. Wolpert}
\email[]{dhw@santafe.edu, web: davidwolpert.weebly.com}% Your name
\affiliation{Santa Fe Institute, Santa Fe, NM, USA}
    \affiliation{Complexity Science Hub, Vienna, Austria}
    \affiliation{Arizona State University, Tempe, AZ, USA}
    \affiliation{International Center for Theoretical Physics, Trieste, Italy}
    \affiliation{Albert Einstein Institute for Advanced Study, New York, NY, USA}

\date{\today}

%TC:ignore
\begin{abstract}

In many complex systems, whether biological or artificial, the thermodynamic costs of communication among their components are large. 
These systems also tend to split information transmitted between any two components across multiple channels. 
A common hypothesis is that such inverse multiplexing strategies reduce total thermodynamic costs. 
So far, however, there have been no physics-based results supporting this hypothesis. 
This gap existed partially because we have lacked a theoretical framework that addresses the interplay of thermodynamics and information in off-equilibrium systems. 
Here we present the first study that rigorously combines such a framework, stochastic thermodynamics, with Shannon information theory. 
We develop a minimal model that captures the fundamental features common to a wide variety of communication systems, and study the relationship between the entropy production of the communication process and the channel capacity, the canonical measure of the communication capability of a channel. 
In contrast to what is assumed in previous works not based on first principles, we show that the entropy production is not always a convex and monotonically increasing function of the channel capacity. However, those two properties are recovered for sufficiently high channel capacity. 
These results clarify when and how to split a single communication stream across multiple channels. 
%In particular, we present Pareto fronts that reveal the trade-off between thermodynamic costs and channel capacity when inverse multiplexing. 
%Due to the generality of our model, our findings could help explain empirical observations of how thermodynamic costs of information transmission make inverse multiplexing energetically favorable in many real-world communication systems. 
    
\end{abstract}

\maketitle

\section{Introduction}

\begin{figure}[htbp]
    \includegraphics[width=0.5\textwidth]{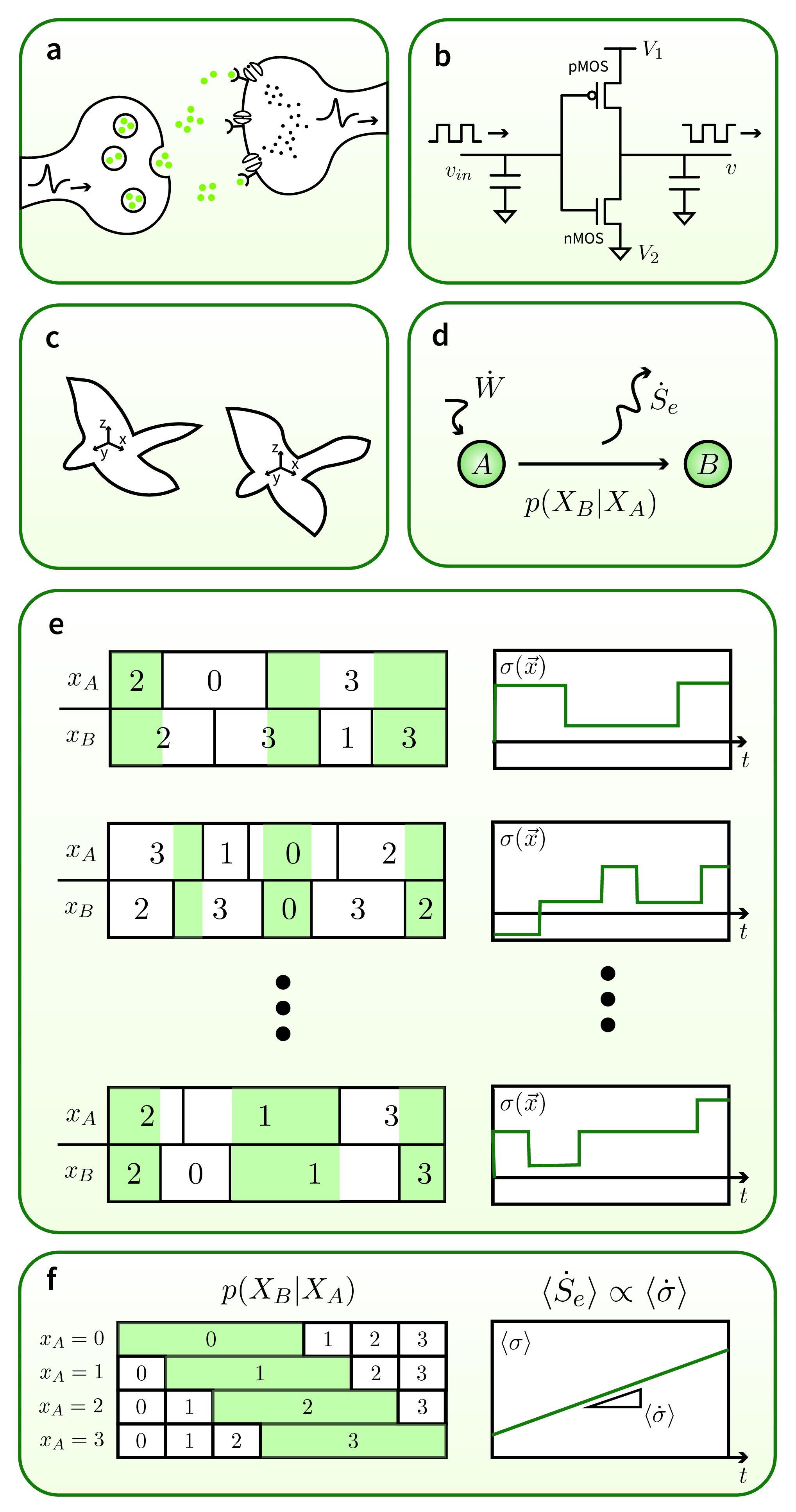}
    \caption{
    \footnotesize
    Examples of communication channels include a) neuronal synapses, b) CMOS inverters, or c) adjacent animals in a flock. These communication channels effectively d) consist of two nodes: an input \textit{A} and an output \textit{B}, which attempts to copy the state of \textit{A}. 
    The conditional probability distribution of the output state given the input state, $p(X_B | X_A)$, reflects the noisiness of the copying process. 
    Since this process must occur out of equilibrium, it requires external work at a rate $\dot{W}$ and dissipates some amount of that work in terms of  entropy flow to the surrounding environment at a rate $\dot{S}_e$. e) Such a communication channel can be modeled as a stochastic process. 
    For a fixed time period of length $t$, the system can have many different realizations, or trajectories $\vec{x}$. 
    These trajectories capture how the state $x_A$ of the input and the state $x_B$ of the output co-evolve. 
    In each trajectory, the output's attempts to copy the state of the input result in an entropy production (EP) $\sigma(\vec{x})$. 
    Boxes highlighted in green represent the time periods for which the state of the output matches the state of the input. 
    Note that there is stochasticity both in the values of the states as well as in the timing of the state transitions. 
    f) Averaged over all possible trajectories, the state occupancies of the input and output define the conditional distribution $p(X_B | X_A)$, and an average EP $\lr{\sigma}$ of the process. 
    Typically the average EP rate, $\lr{\dot{\sigma}}$, accounts for most of the rate of entropy flow to the environment, $\lr{\dot{S}_e}$. 
    } 
    \label{fig:intro}
\end{figure}

One of the major thermodynamic costs of many complex systems arises from communication among their separate subsystems. 
Examples include 
cellular sensing systems~\cite{govern2014cellularsensing, ten2016cellularsensing}, 
the human brain~\cite{balasubramanian2015heterogeneity, balasubramanian2021brainpower},  
% social opinion dynamics~\cite{acemoglu2011opinion, li2019opinionconsensus}, 
ecosystems~\cite{guilford1991ecologicalcommunication, smith2003ecologicalcommunication}, 
wireless sensor networks~\cite{broxton2005localizing, paradiso2005energy, wu2008realistic}, 
hardware implementations of machine learning algorithms~\cite{li2018neuralnetcommunicationcost}, 
and digital computers~\cite{christie2000rentsrule, moses2016energy, demir2014galaxy}. 
Similarly, reducing the thermodynamic costs of communication between processing units and memory units of conventional Von Neumann computational architectures is one of the primary motivations of the field of neuromorphic computing~\cite{prezioso2015hardwareML, markovic2020neuromorphic,  lombo2021neuromorphiccomputing}.
As emphasized in the latest ``Physics of Life'' report from the National Academy of Sciences, it is crucial to understand the \textit{common} physical principles underlying communication in biological systems, whether that be a set of interacting bacteria, a colony of insects, a flock of birds, or a  human social group~\cite{national2022physics}. It is important to appreciate that the information transmission in different communication systems are subject to different physical constraints associated to the transmission process and media (e.g., diffusive, electronic, acoustic, etc. -- see~\cref{fig:intro}(a-c)). In turn, each of these different constraints impose their own, system-specific thermodynamic costs, since they each limit what theoretical efficiencies the system can exploit~\cite{bryant2023physical}. 
However, in order to uncover and investigate the principles common to all types of communication systems, we need a minimal model grounded in the features shared by all of those systems. 

One feature common to all of these systems is that they involve an ``output'' component that is biased to change its state to reflect the state of a separate ``input'' component, which is set exogenously (\cref{fig:intro}(d)). A minimal model that can apply to diverse types of communication systems should include this shared aspect of copying between two separate components. Another feature common to many biological and artificial communication systems is that they often split a stream of information over multiple channels, i.e., they inverse multiplex. Sometimes this occurs even when a single channel could handle the entire communication load. 
For example, multiple synapses tend to connect adjacent neurons~\cite{jones1997induction, meinertzhagen2001synaptic, sorra1993occurrence, toni1999ltp} and multiple neuronal pathways tend to connect brain regions~\cite{fornito2015connectomics, bassett2006small, bassett2017small}. 
In engineering, spatial multiplexing techniques have made multiple-input-multiple-output (MIMO) technology the gold standard for modern wireless communication systems~\cite{gao2019spatial, chen2020beam}. Therefore, analyzing a minimal model of communication should also result in greater insights into the thermodynamics  inverse multiplexing in many scenarios. It may even provide guidance for the design of artificial 
inverse multiplexing systems.

Until now, it has been difficult to construct such a minimal model using a physics-based formalism. The difficulty is that communication systems operate very far from thermodynamic equilibrium, which rules out conventional
arguments based on the quasi-static limit~\cite{shental2008second} or the linear-response regime~\cite{zhou2010minimal}.
% cannot be used to investigate the thermodynamics of communication. 
However, the development of stochastic thermodynamics~\cite{broedersz2022twenty} in the past two decades has supplied a theoretical framework that relates the dynamics of information in a system to the free energy it dissipates in a process arbitrarily far from equilibrium. 
Stochastic thermodynamics can allow us, therefore, to investigate how this thermodynamic cost varies with communication rate in a two-component copying system. That is the goal of this article.

We begin by providing a background on communication theory and stochastic thermodynamics. 
We then build on this background to motivate a minimal model of communication with a well defined thermodynamics. In this simple model, an exogenous agent modulates the energy levels of an output system in order to transmit information. While the output system is in contact with a single heat bath, the switching of its energy levels acts as an external work source that takes it out of equilibrium. Next, we present a preliminary analysis of the thermodynamics of two types of communication encompassed by more general models.
The first one is also a energy switching scenario, but relaxing some of the assumptions made in the simple model mentioned above (namely, we consider several thermal baths and arbitrary relaxation timescales). In the second one, called reservoir switching, the exogenous agent does not change the energy levels of the output system but modulates their coupling to different thermal baths. These two communication types correspond roughly to (i) wireless communication or inter-cellular communication, and (ii) wired electronic communication. 
We show that in many examples of both types of communication, the thermodynamic cost is a convex function of the communication rate above a critical value of that rate. Regardless of the convexity of this function, one might expect that increasing the speed of information transmission through a channel requires a monotonic increase in the thermodynamic costs~\cite{balasubramanian2015heterogeneity, cover1999infotheory}. 
However, we find that in many cases this function is \textit{not} monotonic. 

We then investigate the consequences of this result for when and how one should split a single information stream across multiple physical channels. 
In particular, we derive a Pareto front representing the trade-off between minimizing total thermodynamic cost and maximizing total information transmission rate shared across a fixed number of communication channels.
%This Pareto front can serve as a heuristic for how to distribute a total fixed information capacity across multiple channels in order to minimize the total dissipated work. 
% This analysis may explain why inverse multiplexing arises so often in real-world systems. 
% Since we prove our results for a minimal model of communication in terms of a two-component copying process, these consequences potentially apply to \textit{all} of the examples of real-world communication described above. 
We end by discussing our model in the broader context of the thermodynamics of computation and by suggesting future work.

%TC:ignore	
%TC:endignore

\section*{Background on Shannon information theory and stochastic thermodynamics}

Claude Shannon's channel coding theorem states that one can pass messages through a noisy communication channel with vanishingly low error~\cite{shannon1948commchannel}. 
A channel approaches this error-free limit if the messages sent through it are encoded with an appropriate codebook into strings of letters from a finite, discrete alphabet. 
These encoded strings can be decoded to recover the original message at the output of the channel. 
Importantly, one can implement error-free communication in noisy channels in this way only up to a finite maximum baud rate, called the channel capacity. 
Exceeding that rate imposes a non-zero probability of error. 
A channel's capacity therefore serves as the primary measure of its communication capabilities. 
This capacity equals the maximal mutual information between input and output attainable by varying the input alphabet distribution~\cite{shannon1948commchannel, feinstein1954new}. 

The channel coding theorem further states that one can achieve the channel capacity with negligible error using any one of \textit{many} possible optimal encodings of information at the source\footnote{Although there are countably infinitely many such encodings, it is very difficult for a human to find one~\cite{coffey1990any}.}. 
These encodings produce sequences that look as if they were identically and independently drawn (IID) from the input distribution that maximizes the mutual information between input and output~\cite{slot2015shannon}. 
This theorem led to massive engineering efforts toward developing error-correcting codes that achieve this bound~\cite{berrou1996channelcapacityturbo, chung2001channelcapacityLDPC}. 
Such coding strategies are now widespread in digital communication systems~\cite{duman2008coding, viterbi2013principles, clark2013error, rao2015channel, patil2020coding}. 
Moreoever, the study of codes that approach the Shannon bound has been one of the major areas of research in modern information theory for several decades. 

The duality between information and Shannon entropy creates a tight link between information theory and thermodynamics~\cite{brillouin1951maxwell, landauer1961irreversibility, esposito2011second}. 
% Recent discoveries in stochastic thermodynamics have expanded the inviolability of the Second Law at the macroscopic scale is matched by a complete absence of the Second Law at the microscopic scale. 
Stochastic thermodynamics has revealed 
that the entropy produced over any finite time interval in a single realization of a nonequilibrium process\footnote{In particular, stochastic thermodynamics describes any nonequilibrium system whose state transitions are mediated by energy exchanges with infinite reservoirs.} \textit{can} be negative, in defiance of the Second Law. 
This entropy production (EP) recovers non-negativity only when averaged over all possible realizations (trajectories)~\cite{vdbesposito2015ensemble}\footnote{Moreoever, as the number of degrees of freedom in the model of the system increases, the probability that the entropy production is negative decreases~\cite{rao2018DFT}. This explains the observation that the inviolability of the Second Law at the macroscopic scale is matched by a complete absence of the Second Law at the microscopic scale~\cite{collin2005verification, joubaud2008fluctuations}.}. 
% , a fact that went completely unnoticed in the 20th century.
Throughout the paper, we refer to the EP as the thermodynamic cost. 
This measure is crucial in many powerful results, including bounds on the precision of generalized currents~\cite{barato2015TUR, horowitz2017TUR} and the speed of changes in a system's probability distribution over states~\cite{shiraishi_funo_saito_2018}. 
The EP has also been applied to study the dynamics of information~\cite{parrondo2015thermodynamics}. 

Unfortunately, most of the prior work in stochastic thermodynamics involving mutual information and ``information processing'' fails to relate the 
%framework's primary measure of the dissipated work, the entropy production (EP), 
EP to the channel capacity.  
%despite the clear theoretical and practical importance of the channel capacity. 
In fact, most of these results do not consider a system dedicated to communication. 
% formalized as implementing the identity map. 
Some studies analyze how components influence one another in multipartite systems of multiple feedback controllers by quantifying the ``information flow'' between subsystems~\cite{horowitz2014bipartiteinfoflow, horowitz2015multipartiteinfoflow, sagawa2012feedback, sagawa2012thermodynamics}. 
Other studies, e.g., on Bayes' Nets~\cite{ito2013information}, use the ``transfer entropy'' rather than the channel capacity to characterize information transferred from inputs to outputs. 
Studies of specific models for biomolecular copying processes~\cite{poulton2019nonequilibrium} and cellular sensing~\cite{ito2015maxwell, hartich2016sensory, bryant2023physical} have made analogies with information theory and the channel coding theorem, but are limited in scope to only biological systems. 
Similarly, the analysis of channel capacity in~\cite{gao2021circuits} applies only to the very restricted domain of electronic circuits comprising a set of transistors running sub-threshold. 

An early connection between channel capacity and thermodynamic cost appeared naturally in the study of continuous Gaussian channels. These channels are an idealization of the information transmission through a telegraphic/telephonic line or radio link. Without additional constraints such continuous channels have infinite capacity, which is not physical. A natural and realistic constraint is to impose a maximum power on the physical signal used to transmit information. Since the medium through which the signal propagates is linear, the power is just a quadratic function of the signal intensity. Under the previous assumptions, it was understood early on that the capacity $C$ of a Gaussian channel is related to its bandwidth $W$, its noise spectral density $N_0$ and the power $P$ of the input signal as \cite{cover1999infotheory}
\eq{
C = W \ln\left(1+\frac{P}{N_0W}\right).
\label{eq:gaussian}
}
In the infinite bandwidth limit we have $C \simeq P/N_0$. Also, for a medium in thermal equilibrium at temperature $T$ the power spectral density of the noise is $N_0 = k_b T$, which implies that one must transmit a signal with an energy of at least $E = n k_bT$ in order to communicate $n$ bits. However, one must note that the input power $P$ is not necessarily dissipated in the transmission medium, and cannot be directly related with the entropy produced in the transmission process. Therefore, until now the literature has lacked an investigation of the quantitative relationship between the channel capacity and the EP that could apply to all types of communication, even those with a finite set of discrete symbols. We conduct exactly this investigation, justifying the features of our minimal model with the rigor of Shannon information theory. As a final comment, note that according to Eq. \eqref{eq:gaussian} the transmitted power $P$ is an increasing convex function of the channel capacity $C$. Those two properties, the monoticity and the convexity of the thermodynamic cost for any $C$, are sometimes assumed in some works that are not based on first principles \cite{balasubramanian2001metabolically}. Interestingly, as we will see, they are not preserved in general models of communication channels with discrete set of symbols. They are however recovered for sufficiently large channel capacity.

\section*{Stochastic thermodynamics of communication channels}

We analyze the thermodynamics of communication channels operating at their channel capacity. 
In Shannon's model of communication, a process external to the system sets the input state so that its dynamics reflect samplings from the input distribution that achieves the channel capacity. 
This external system effectively acts as a work reservoir, and could follow arbitrary (potentially non-Markovian) dynamics. 
The input could be set deterministically or stochastically, periodically according to a clock, according to a continuous-time Markov chain (CTMC), etc. 
%Accordingly, we assume that the input has no explicit reservoirs that mediate its state transitions. 
This feature is physically motivated by the fact that in many real communication channels  the input is set by an exogenous process. 
Therefore, we ignore the thermodynamics of the external system that sets the input. 
% , meaning that we consider the thermodynamics of a system that copies the state of an external environment. 
% This relates our investigations to previous studies in the thermodynamics of copying, observation, and sensing~\cite{hartich_barato_seifert_2016, sagawa2009minimal}. 
Finally, the coupling between input and output is non-reciprocal, in that the output dynamics depends on the state of the input but not vice-versa. 
This common assumption is sometimes called ``no back-action''~\cite{sagawa2009minimal, sagawa2012feedback, hartich_barato_seifert_2016, verley2014nobackaction, bryant2020energy, bryant2023physical, wolpert_min_ep_2020}. 
% This property reflects the fact that in real communication systems, the thermodynamics of setting the value of the input is not provided by the communication device that then transports that information downstream~\cite{}. 

We consider a communication channel in which the input $A$ sends letters $x_A$ from the alphabet $X_A = \{1, 2, \ldots, L \}$ in any way that reflects the distribution $\pi_{X_A}$. 
%This distribution over input states maximizes the mutual information between input and output, i.e., achieves the channel capacity. 
The communication channel is defined by how the output $B$ changes its state, $x_B \in X_B = \{1, 2, \ldots, L \}$, in response to the input state.
% Based on the state of the input, the output changes its state $x_B \in X_B = \{0, 1, \ldots, L \}$. 
At any given time, the rates of state transitions in the output depend on the current state of the input. 
The energies of different output states may depend on the state of the input according to a Hamiltonian function $H(x_B|x_A)$. In addition, the coupling of the output system to different thermodynamic reservoirs may also depend on the state of the input A. 

\subsection{A minimax bound on the channel capacity}

To begin, we derive a general bound on the capacity of a channel that will be useful for the subsequent thermodynamic analysis. The joint distribution for input-output pairs can be written as $p_{x_B, x_A} = p_{x_B|x_A} \pi_{x_A}$ in terms of the conditional distribution $p_{x_B|x_A}$. The mutual information between input and output can be written as:
\begin{equation}
    I(A:B) = \sum_{x_A} \KLD{p_{x_B|x_A}}{p_{x_B}}\: \pi_{x_A},
\end{equation}
where $p_{x_B} = \sum_{x_B} p_{x_B, x_A}$ is the unconditional distribution over outputs and $\KLD{p}{q}$ is the Kullback-Leibler divergence or relative entropy between distributions $p$ and $q$.
The channel capacity is defined as the maximum of $I(B:A)$ over all possible input distributions $\pi_{x_A}$. Thus, we do a unconstrained optimization of the function $G \equiv I(B:A) - \alpha (\sum_{x_A} \pi_{x_A} - 1)$ over $\{\pi_{x_A} \geq 0\}$ and the Lagrange multiplier $\alpha$. After some algebra, we find:
\begin{equation}
    \frac{\partial G}{\partial \pi_{x_A}} = 
    -1-\alpha + D(p_{x_B|x_A} || p_{x_B}).
\end{equation}
Then, for the optimal solution satisfying 
$\partial G/\partial \pi_{x_A} = 0$, the relative entropy $D \equiv  D(p_{x_B|x_A} || p_{x_B})$ must be independent of $x_B$. Also, in that case we
have $I(A:B)=D$. Therefore, the channel capacity is $C=D$.

This analysis does not provide an explicit expression for the channel capacity, since that would require finding the distribution ${\pi_{x_A}}$ solving $\partial G/\partial \pi_{x_A} = 0$.
However, it allows to obtain a useful upper bound. Indeed, for any $x_A$, we have:
\begin{equation}
\begin{split}
C = D(p_{x_{B}|x_{A}}||p_{x_B}) &= \sum_{x_B} p_{x_B|x_A} 
\log\left( \frac{p_{x_B|x_A}}{p_{x_B}} \right)\\
&\leq \sum_{x'_A,x_B} \pi_{x'_A} \: p_{x_B|x'_A} 
\log\left( \frac{p_{x_B|x_A}}{p_{x_B|x'_A}} \right) \\
& = \sum_{x'_A} \pi_{x'_A} \KLD{p_{x_B|x_A}}{p_{x_B|x'_A}},
\end{split}
\label{eq:bound_capacity}
\end{equation}
where in the second line we used Jensen's inequality. It follows that 
\begin{equation}
    C \leq \max_{x'_A} \KLD{p_{x_B|x_A}}{p_{x_B|x'_A}}
\end{equation}
for any $x_A$. Then 
\begin{equation}
    C \leq \min_{x_A} \max_{x'_A} \KLD{p_{x_B|x_A}}{p_{x_B|x'_A}}.
    \label{eq:minimax_bound}
\end{equation}

The meaning of this inequality is clear if we interpret the relative entropy $\KLD{p_{x_B|x_A}}{p_{x_B|x'_A}}$ as a distance between those conditional output distributions. Indeed, if that distance is small, it means that the output distribution does not change too much for different inputs, and therefore the mutual information between input and output must also be small. 
%Also, the previous bound has a thermodynamic meaning derived from the fact that the relative entropy between an arbitrary distribution and a thermal equilibrium distribution gives the average free energy dissipated in a relaxation process towards the corresponding equilibrium \cite{esposito2011second}, as we exploit in the following.

\subsection{Energy switching communication: the case of a single reservoir and timescale separation}

Before studying the problem in full generality, we examine what probably is the simplest model of communication channel with a well defined thermodynamics. As we will see, this model is already non-trivial, and displays many features shared by more general settings examined later. 

In this model, the output system $B$ is in contact with a single thermal reservoir at temperature $T$, and the only effect of the input $A$ is to control the energies of the output states according to the Hamiltonian:
\eq{
    H(x_B| x_A) = 
    \begin{cases}
        -\epsilon & x_B = x_A\\
        0 & \text{otherwise}
    \end{cases}.
    \label{eq:hamiltonian}
}
Thus, the output state matching the input state is energetically favoured.
Eq. \eqref{eq:hamiltonian} can also be understood as the time-independent Hamiltonian of the global system.
For a constant input
$x_A$, the steady state of the output is given by the equilibrium distribution:
\eq{
    p_{x_B|x_A} = 
    \begin{cases}
        \frac{e^{\beta\epsilon}}{e^{\beta\epsilon} + L-1} & x_B = x_A\\
        \frac{1}{e^{\beta\epsilon} + L-1} & \text{otherwise}
    \end{cases},
    \label{eq:simple_channel}
}
where $\beta=1/T$ (we take $k_b=1$). Recall that $L$ is the number of distinct states in both the input and output systems. The average energy and entropy of the output are therefore independent of the input and read, respectively
\eq{ 
U = -\epsilon \frac{e^{\beta\epsilon}}{e^{\beta\epsilon} + L-1} \quad \text{and} \quad
S = \ln(e^{\beta\epsilon} + L-1) + \beta U.
\label{eq:simple_energy_entopy}
}
Also, we consider that the waiting times between successive changes of the input are much longer than the relaxation time of the output. Under such time-scale separation assumption, the entropy produced during the relaxation of the output following a change $x_A \to x_A'$ in the input is given by the relative entropy between the initial and the final distributions \cite{esposito2011second}. Thus, we have:
\eq{
\sigma_{x_A \to x_A'} \equiv
\KLD{p_{x_B|x_{A}}}{p_{x_B|x'_A}} = 
-\beta \frac{\epsilon + LU}{L-1}
%\beta\epsilon \frac{e^{\beta\epsilon}-1}{e^{\beta\epsilon} + L-1},
\label{eq:simple_ep}
}
Since the previous equation is actually independent of the input transition, the average rate of entropy production is just $\dot \sigma = f_s \sigma$, where $f_s$ is the average frequency of input transitions. Note also that Eq. \eqref{eq:simple_ep} and the minimax bound of Eq. \eqref{eq:minimax_bound}
allow to relate the entropy production with the channel capacity. This will be used later to define the thermodynamic efficiency of the channel.

In this symmetric case the channel capacity is achieved for a uniform input distribution and simply reads:
\eq{
C = \ln(L) - S,
}
where $S$ is the entropy of the output system given in Eq. \eqref{eq:simple_energy_entopy}.

We want to study the relationship between the channel capacity and the entropy production. From the previous equations it follows that:
\eq{
\frac{d\sigma}{dC}  = -\frac{d\sigma}{dS} = \frac{1}{L-1} \left[ L + \frac{\epsilon + LU}{\beta U(\epsilon + U)}\right].
}
Note that $d\sigma/dC \geq 0$, since $U \leq -\epsilon/L$. Then, the entropy production increases with the channel capacity. 

After some algebra, the second derivative of the entropy production with respect to the channel capacity can be found to be
\eq{
\frac{d^2\sigma}{d^2C} = 
\frac{1}{L-1} \frac{1}{\beta^2 U^2(\epsilon + U)}
\left[\epsilon + (\beta^{-1} + U)\frac{\epsilon + LU}{\epsilon+U}\right].
}
It can be seen that $d^2\sigma/d^2C \geq 0$
if and only if 
\eq{
e^{2\beta \epsilon} (\beta\epsilon -1) - e^{\beta\epsilon} (L-2) + (\beta\epsilon +1)(L-1) \geq 0,
}
which is always true for $L=2$. For a given $L>2$, we have $d^2\sigma/d^2C < 0$ for small $\beta\epsilon$ and $d^2\sigma/d^2C > 0$ for large $\beta\epsilon$. As a consequence, as a function of the channel capacity $C$, the entropy production $\sigma$ is concave for small $C$ and convex above a critical value.

Finally, we note that in the limit $\beta \epsilon \gg 1$ in which $C$ approaches $\ln(L)$, the entropy production diverges as
\eq{
\sigma \simeq \ln\left(\frac{L-1}{\ln(L) - C}\right) + \ln\left(1+\ln\left(\frac{L-1}{\ln(L) - C}\right)\right).
}

\begin{figure}
    \centering
    \includegraphics[scale=.6]{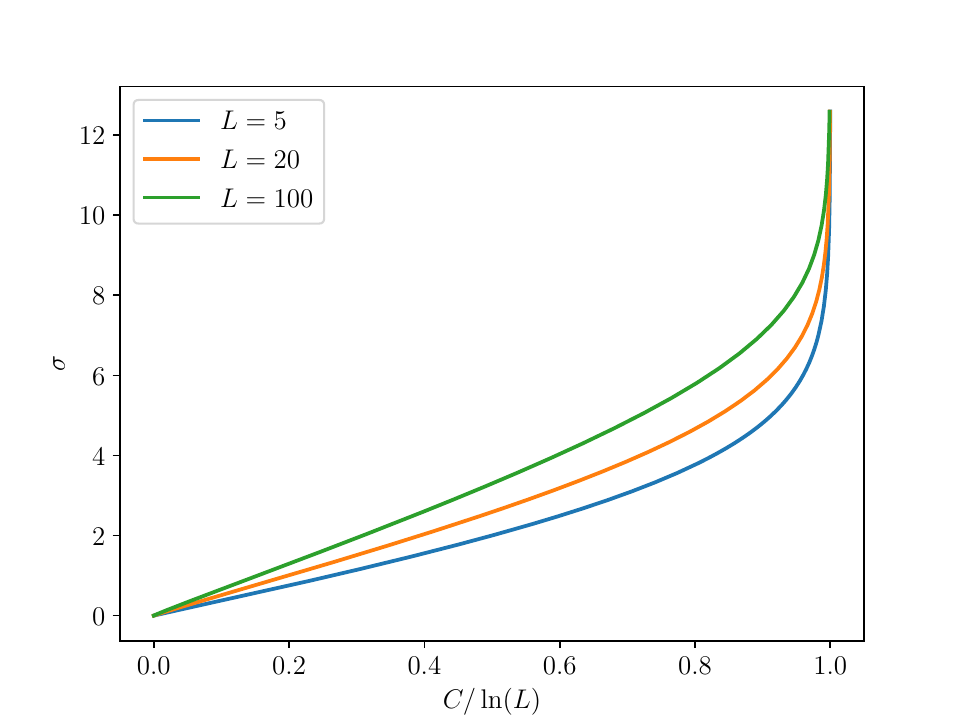}
    \caption{Entropy production versus channel capacity for the channel given in Eq. \eqref{eq:simple_channel}.}
    \label{fig:simple_ep_vs_c}
\end{figure}

Figure \ref{fig:simple_ep_vs_c} shows the entropy production as a function of the channel capacity. The concave region at low capacity is too flat and too small to notice in the plot. 

\begin{figure}
    \centering
    \includegraphics[scale=.6]{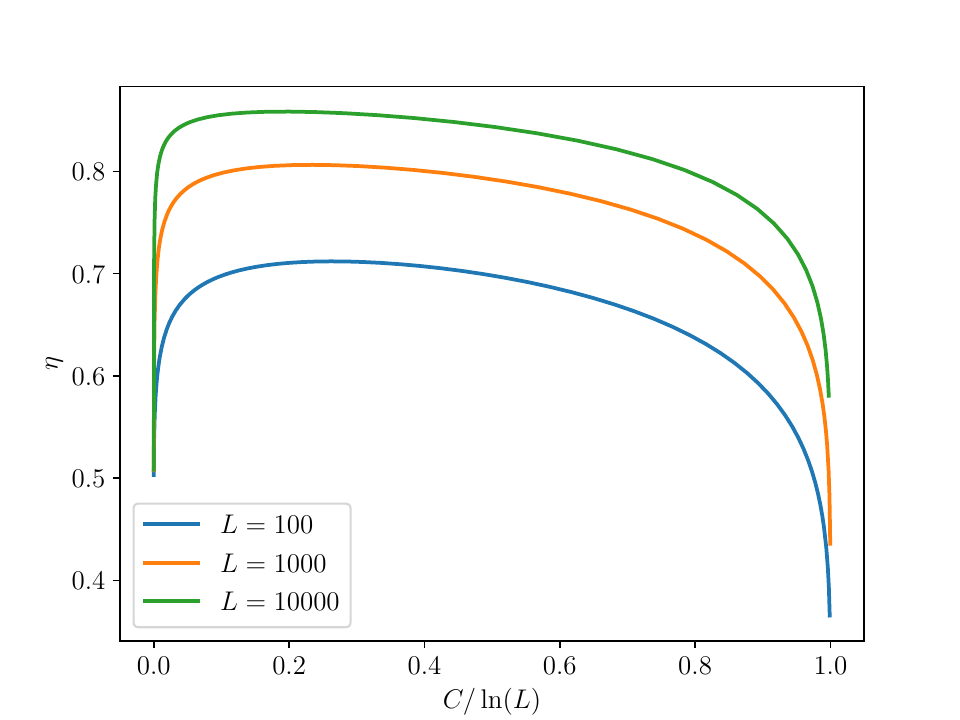}
    \caption{Efficiency versus channel capacity for the channel given in Eq. \eqref{eq:simple_channel}.}
    \label{fig:simple_eta_vs_c}
\end{figure}

Another interesting quantity to study is the ratio $\eta \equiv C/\sigma$ between channel capacity and entropy production, which can be considered the thermodynamic efficiency of the channel. It is shown in Figure \ref{fig:simple_eta_vs_c}, where we see that we always have $\eta \leq 1$. That this must be the case is a direct consequence of Eqs. \eqref{eq:minimax_bound} and \eqref{eq:simple_ep}.

\subsection{General case}

We now study more general models where we relax the main assumptions made above, namely i) that the output system is in contact with a single thermal bath, ii) that its relaxation time is shorter than the typical time between input transitions, and iii) that the only effect of the input is to modulate the energies of the output states. 

The dynamics of both input and output states is modeled as a CTMC (\cref{fig:intro}(e,f)). Then, the joint distribution $p_{x_A,x_B}(t)$  evolves according to the master equation:
\eq{
	\dot{p}_{x_A,x_B}(t) = 
    \sum_{x'_A, x'_B} R^{x'_A,x'_B}_{x_A,x_B} \: p_{x'_A,x'_B}(t)
}
where $R^{x'_A,x'_B}_{x_A,x_B}$ are the elements of a $L^2 \times L^2$ rate matrix. Each non-diagonal element of that matrix indicates the probability of observing a state transition $(x'_A,x'_B) \to (x_A,x_B)$ at any given time. The diagonasl elements are such that the normalization is preserved.

The input and output together evolve as a bipartite process\footnote{A bipartite process is a special case of a multipartite process (MPP)~\cite{wolpert_min_ep_2020,horowitz2014bipartiteinfoflow} involving only two subsystems.}, so only one subsystem changes state at any given time. Thus, each non-diagonal element of the rate matrix obeys
\eq{
    R^{x'_A,x'_B}_{x_A,x_B} &= \delta^{x'_B}_{x_B} \, J^{x'_A}_{x_A}(x_B)  + \delta^{x'_A}_{x_A} \, K^{x'_B}_{x_B} (x_A),
}
where $\delta^{x'}_{x}$ is the Kronecker delta function that equals $1$ when $x' = x$, and equals $0$ otherwise.
Additionally, the output $B$ changes its state due to interactions with a set of $N$ equilibrium reservoirs $\V := \{ v_1, v_2, \ldots, v_N \}$, at temperatures $\{ T_1, T_2, \ldots, T_N \}$. We denote the temperature of reservoir $v$ as $T_v$. 
This means that the overall rate matrix  $K^{x'_B}_{x_B} (x_A)$ can be written as a sum over the rate matrices representing each reservoir's effect on the output's state transitions:
\eq{
     K^{x'_B}_{x_B}(x_A) = \sum_{\nu = 1}^N K^{x'_B}_{x_B}(x_A;v).
}
Each rate matrix specific to reservoir $v$ follows local-detailed balance:
\eq{
	\ln \frac{K^{x'_B}_{x_B}(x_A; v)}{K_{x'_B}^{x_B}(x_A; v)} = -\frac{1}{T_v} \left( H(x_B|x_A) - H(x'_B|x_A) \right).
    \label{eq:ldb}
}

\noindent
The previous condition allows to decompose each matrix $K^{x'_B}_{x_B}(x_A; v)$ as the product of a diagonal matrix encoding the equilibrium distribution corresponding to temperature $T_v$, and a symmetric matrix that contains the information about how the output system is coupled to that environment (and thus determines how the relaxation to the previous equilibrium takes place). Indeed one can write the non-diagonal elements  
$K^{x'_B}_{x_B}(x_A; v)$ as
\eq{ \label{R-param}
K^{x'_B}_{x_B}(x_A; v)
= \sum_{x''_B}   \pi_{x_B}(x_A;v) \delta^{x''_B}_{x_B}
\: S^{x'_B}_{x''_B}(x_A; v) 
}

\noindent
where $S^{x'_B}_{x_B}(x_A; v)$ are the elements of a symmetric matrix and $\pi_{x_B}(x_A; v) = e^{-H(x_B | x_A)/T_v}/Z$, with $Z = \sum_{x_B} e^{-H(x_B | x_A)/T_v}$, is the equilibrium output distribution corresponding to reservoir $v$. Note that although every reservoir-specific rate matrix is detailed-balanced, the overall rate matrix is not detailed-balanced in general unless there is only a single reservoir or all of them are at the same temperature.

The bipartite nature of the overall system's evolution implies that the rate of entropy production $\dot \sigma$ can be split as~\cite{horowitz2014bipartiteinfoflow, wolpert_min_ep_2020}:
\begin{equation}
    \dot \sigma = \dot \sigma_A + \dot \sigma_B.
\end{equation}
Whereas $\dot{\sigma}_A$ represents the cost of switching the input, $\dot{\sigma}_B$ represents the thermodynamic cost of the copying process. At any given time, the later can be computed as:
\eq{ 
    \spl{
        \dot{\sigma}_B &= %\\
        \sum_{\substack{v,x_A, \\ x'_B,x_B}} K^{x'_B}_{x_B}(x_A;v) p_{x_A,x'_B}(t) \: \ln \frac{K^{x'_B}_{x_B}(x_A;v) p_{x_A,x'_B}(t)}{K^{x_B}_{x'_B}(x_A;v) p_{x_A,x_B}(t)} 
    }
}
A similar expression holds for $\dot \sigma_A$.

Within this minimal model, we analyze two general methods that can modulate the output dynamics in real communication systems. 
In the ``energy switching'' case, different values of the input modulate the energies of different output values, as in the simple case above. 
Intercellular chemical communication in biological organisms can be modeled in this way. 
For example in cellular sensing, the concentration of a ligand (input) modulates the free energy of receptor binding (output)~\cite{hartich_barato_seifert_2016}\footnote{The fraction of bound receptors on the cellular membrane reflects the external ligand concentration.}. 
Additionally, we can model wireless communication in terms of energy switching, where electromagnetic waves sent by the transmitter modify the potential energy function of electrons at the receiver. 
In the ``reservoir switching'' case, different values of the input modulate the coupling of the output to different reservoirs. 
We can model wired electronic communication in circuits in this way, where the input voltage modulates the output wire's couplings with different chemical reservoirs of electrons~\cite{freitas2021stochthermocircuits}. 

\subsection{Energy switching scenario}

In the energy switching method of communication, the energy levels of different output states depend on the input state. 
We analyze the class of channels for which the output achieves its lowest energy state when it matches the current state of the input, and all other ``mismatched states'' have the same higher value of energy. Thus the Hamiltonian is also given by Eq. \eqref{eq:hamiltonian}.
As before, since the Hamiltonian is symmetric with respect to every letter in the input alphabet, the channel capacity-achieving input distribution $\pi_{X_A}$ is the uniform distribution ($\pi_{x_A} = 1/L$). Additionally, we set all non-zero off-diagonal entries of the matrix $S(x_A; v)$ to a constant $r_v$, so that the couplings of the output to its different reservoirs cannot be modulated by the state of the input. 
Therefore, every non-zero off-diagonal rate matrix element equals
\eq{
    K^{x'_B}_{x_B}(x_A;v) = 
    \begin{cases}
        r_v \frac{e^{\epsilon/T_v}}{L(e^{\epsilon/T_v} + (L-1))} =: \alpha_i & x'_B \neq x_A ; x_B = x_A \\
        r_v \frac{1}{L(e^{\epsilon/T_v} + (L-1))} =: \zeta_i & x_B \neq x_A 
    \end{cases}
}
In our simulations for the energy switching case, the input dynamics follow a telegraph process~\cite{meijers2021infoflow}, which is consistent with a CTMC. 
A telegraph process serves as an accurate model for any real communication channel (e.g., synaptic release~\cite{markovic2020physics}) for which the waiting times for the input in any given state follows an exponential distribution. 
%Examples of systems with such input dynamics
The input switches its state with average rate $f_s$, which is a positive real number.  
This means that each state transition in the input is equally likely, so that $R^{x'_A,x_B}_{x_A,x_B} = J^{x'_A}_{x_A}(x_B) = f_s$, for $x'_A \neq x_A$. 
Therefore, the non-zero off-diagonal elements of the overall channel's rate matrix are

\eq{
    R^{x'_A,x'_B}_{x_A,x_B} = 
    \begin{cases}
        f_s & x'_A \neq x_A; x'_B = x_B \\
        \alpha := \sum_{i=1}^N \alpha_i & x'_A = x_A ; x'_B \neq x_A ; x_B = x_A  \\
        \zeta := \sum_{i=1}^N \zeta_i  & x'_A = x_A ; x_B \neq x_A 
    \end{cases}
}

We first analyze the situation where the channel is in a nonequilibrium steady-state (NESS), which means
\eq{
    \sum_{x'_A,x'_B} R^{x'_A,x'_B}_{x_A,x_B} \pi_{x'_A,x'_B} = 0
}
Solving for the joint steady-state distribution $\pi_{X_A,X_B}$, we obtain  

\eq{
    \pi_{x_A,x_B} = 
    \begin{cases}
        \frac{\alpha + f_s}{L\left[\alpha + (L-1)\zeta + Lf_s \right]} := \frac{p_m}{L} & x_A = x_B  \\
        \frac{\zeta + f_s}{L\left[\alpha + (L-1)\zeta + Lf_s \right]} := \frac{p_e}{L} & \text{otherwise}
    \end{cases}
}

\noindent
This means that the conditional steady-state distribution of the output state given the input state reads

\eq{
    \pi_{x_B|x_A} = 
    \begin{cases}
        p_m & x_B = x_A \\
        p_e & \text{otherwise}
    \end{cases}
}

\noindent
and the marginal steady-state distribution over just the output states is uniform. Note that the previous conditional distribution differs from an equilibrium distribution even in the case there is a single thermal reservoir, except in the limit $f_s \to 0$. The reason is that the changing input plays the role of a work reservoir that parametrically changes the energy landscape of the output. In the opposite limit $f_s \to \infty$ the conditional distribution is just uniform, since the output cannot adapt to the rapidly changing input.

% We find that the EP rate in the NESS is
% \eq{ 
%     \spl{
%         \lr{\dot{\sigma}} &= %\\
%         \sum_{\substack{v,x_A, \\ x'_B,x_B}} K^{x'_B,x_A}_{x_B,x_A}(B;v) \pi_{x'_B,x_A} \ln \frac{K^{x'_B,x_A}_{x_B,x_A}(B;v) \pi_{x'_B,x_A}}{K_{x'_B,x_A}^{x_B,x_A}(B;v) \pi_{x_B,x_A}} \\
%         &+ \sum_{x_B,x'_A, x_A} K^{x_B, x'_A}_{x_B,x_A}(A) \pi_{x_B,x'_A} \ln \frac{K^{x_B,x'_A}_{x_B,x_A}(A) \pi_{x_B,x'_A}}{K_{x_B,x'_A}^{x_B,x_A}(A) \pi_{x_B,x_A}} 
%     } \label{epr-gen}
%      \\
%     \spl{
%         &= \sum_{i=1}^N (L-1) \left(\zeta_i p_m - \alpha_i p_e \right) \ln \frac{\zeta_i p_m}{\alpha_i p_e} \\
%         &\qquad+ (L-1)f_s \left( p_m - p_e \right) \ln \frac{p_m}{p_e}
%     } \label{epr-es} \\
%     &= \lr{\dot{\sigma}_B} + \lr{\dot{\sigma}_A}
% }
% \noindent
% where the EP rate due only to the output (input) transitions, $\lr{\dot{\sigma}_B}$ ($\lr{\dot{\sigma}_A}$), is equal to the top (bottom) line in each of \cref{epr-gen} and \cref{epr-es}.
% %and the EP rate due only to the input transitions, $\lr{\dot{\sigma}_A}$, is equal to the bottom line in each of \cref{epr-gen} and \cref{epr-es}. 

%TC:ignore	
\begin{figure*}[htbp]
	\includegraphics[width=\textwidth]{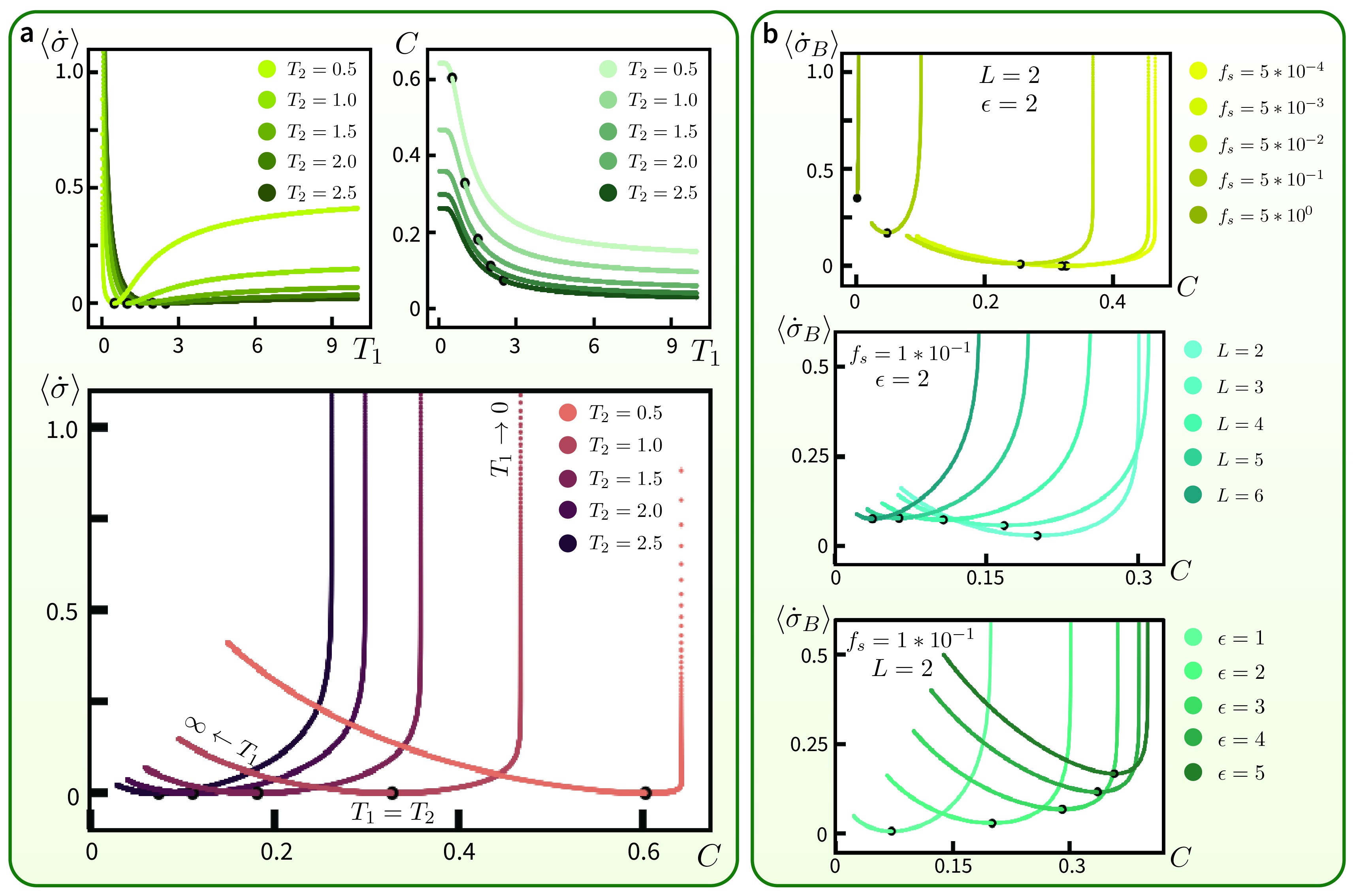}
	\caption{
        \footnotesize
		Results for the case when the state energies of the communication channel's output vary with the input state. 
        Here the output couples to two reservoirs with temperatures $T_1$ and $T_2$. 
        We vary the noise in the channel by holding the temperature of one reservoir fixed and varying the temperature of the other reservoir. 
        Black dots in each graph indicate $T_1 = T_2$. 
        \textbf{(a)} These plots reflect the case of a binary alphabet ($L=2$) and for energy bias $\epsilon = 2$. 
		\textbf{(a, top left)} We first analyze the case of zero input signal rate: $f_s = 0$. 
        We find that the EP rate $\lr{\dot{\sigma}} = \lr{\dot{\sigma}_B}$ as a function of $T_1$ has a single global root when $T_1 = T_2$. 
        The derivative of the EP rate is negative (non-negative) to the left (right) of this root. 
        \textbf{(a, top right)} The channel capacity $C$ is a positive function of $T_1$ and its derivative is non-positive. 
		\textbf{(a, bottom)} Combining these two relationships, we find that the EP rate has at most one minimum with respect to the channel capacity. 
        As a result, $\lr{\dot{\sigma}}(C)$ is a convex and non-monotonic function (see proof in Materials and Methods).
		\textbf{(b)} Plots of the function $\lr{\dot{\sigma}_B}(C)$ for \textbf{(top)} different signaling rates $f_s > 0$, 
        % (Note that when the signaling rate is too high, the channel capacity is almost always zero.)
        \textbf{(middle)} different alphabet lengths $L$, and \textbf{(bottom)} different energy biases $\epsilon$. 
		We observe in all plots that the EP rate retains a single global minimum (marked by black dots) with respect to the channel capacity. 
	} \label{fig:results}
\end{figure*}
%TC:endignore

Since the input distribution is chosen so that the system runs at its channel capacity, the value of the channel capacity in the NESS is simply given by the mutual information between the input and the output:

\eq{
    C &= I(A:B) = S(B) - S(B|A) \\
    &= \ln L + p_m \ln p_m + (L-1) p_e \ln p_e.
}

\noindent
%where the entropy of the output distribution is $S^{X_B} = \sum_{x_B} - \pi_{x_B} \ln \pi_{x_B} $ and the entropy of the conditional distribution of the output given the input is $S^{X_B|X_A} = \sum_{x_B, x_A} - \pi_{x_B|x_A}\pi_{x_A} \ln \pi_{x_B|x_A} $. 

\cref{fig:results} shows results of analyzing how the thermodynamic cost varies with the channel capacity when we adjust the amount of noise in the channel. 
We restrict attention to the case when the system is out of equilibrium due to the presence of two thermal reservoirs at different temperatures ($N = 2$).
To adjust the noise level in the channel, we sweep the temperature of one reservoir $T_1$ while holding fixed the temperature of the other at $T_2$. 
We find that $\lr{\dot{\sigma}_B}$ as a function of $C$ is convex with a single global minimum that occurs at a positive value of the channel capacity. 
In particular, this means that the thermodynamic cost is not a monotonic function of the channel capacity. 
We prove this convex, non-monotonic relationship in the limit of low signaling rate (Materials and Methods).

\subsection{Reservoir switching scenario}

The channel can alternatively modulate the coupling of the output system with its different reservoirs.
For this reservoir switching case, the energies  $H(x_B | x_A) = H(x_B)$ of the output states do not depend on the input. 
So, the equilibrium distributions  $\pi_{x_B}(x_A;v)$ corresponding to each reservoir are independent of the input state. Then, for communication to be possible, the components of the symmetric matrix $S^{x'_B}_{x''_B}(x_A; v)$ in \cref{R-param} must be input dependent.  

% Consider an example based on the formalism of the previous section. Let $N = 2$, $L=2$, and $\Delta E = H(x_B=2) - H(x_B=1) > 0$. 
% Set the temperatures $T_1 \ll \Delta E$ and $T_2 \gg \Delta E$ and modulate the reservoir couplings according to the input state:

% \eq{
%     \spl{
%         x_A=1: \qquad r_{v_1}(1) \gg r_{v_2}(1) \\
%         x_A=2: \qquad r_{v_1}(2) \ll r_{v_2}(2) 
%     }
% }

% \noindent
% Then, when the input $x_A=1$ the output's interaction with the low-temperature reservoir $v_1$ means the output spends most of its time in the ground state $x_B=1$. 
% When the input $x_A=2$, the output's interaction with the high-temperature reservoir $v_2$ biases the output to fluctuate often and spend approximately the same amount of time in both of its possible states. 
% In this way, changes in the reservoir couplings correlate the output state with the input state.

Importantly, this kind of communication protocol is used to transfer information between different components of modern electronic circuits. 
For example, consider a CMOS inverter (\cref{fig:intro}(b)), which is the electronic implementation of a NOT gate.
The output voltage $v$ of the inverter connects to two chemical reservoirs of electrons at fixed voltages $V_1$ and $V_2$ (with $V_1>V_2$), through two complementary MOS transistors (a nMOS and a pMOS). 
% The input voltage $v_{\text{in}}$ controls the conductivity of these transistors. 
Increasing $v_{\text{in}}$ increases the conductivity of the nMOS transistor and decreases the conductivity of the pMOS transistor. 
So for high input voltage, the output's interaction with the reservoir at voltage $V_2$ dominates, making $v\simeq V_2$ at steady state. 
The situation reverses for low input voltages, resulting in $v \simeq V_1$. However, the energy associated to an output voltage $v$ is always $E(v) = C_o v^2/2$ ($C_o$ is the output capacitance), independently of the input.

We consider the following minimal model of the reservoir switching case. The input can fluctuate between $L=2$ states. Transition $x_A: 0 \to 1$ has rate $f_+$ and the reverse transition has rate $f_-$.
The output has a single ground state with zero energy, and $M$ degenerate excited states with energy $\epsilon$. Over that set of microscopic states, we define two macrostates $x_B=0$ or $x_B=1$ corresponding respectively to the ground state being occupied, or any of the excited states being occupied. The output is in contact with two thermal reservoirs at temperatures $T_v$ ($v=1,2$) that can induce transitions between the ground and excited microstates but not between the excited microstates themselves. In this case, the coarse grained rates between macrostates satisfy the local detailed balance conditions \cite{esposito2012CG}:
\begin{equation}
    \ln \frac{\lambda_+(x_A; v)}{\lambda_-(x_A;\nu)} = -\frac{1}{T_v}(\epsilon - T_v \ln(M)),
    \label{eq:ldb1}
\end{equation}
where $\lambda_+(x_A;v)$ is the rate of the transition $x_B: 0 \to 1$ induced by the reservoir $v$ for input $x_A$, and $\lambda_-(x_A; v)$ is the rate of the reverse transition. The term $\ln(M)$ takes into account the internal entropy of the macrostate $x_B=1$. The equilibrium distributions $\pi_{x_B}(x_A;v)$ are then:
\begin{equation}
    \pi_{x_B}(x_A;v) = 
    \begin{cases}
    \frac{M e^{-\epsilon/T_v}}{1+M e^{-\epsilon/T_v}} 
    & \quad  x_B = 1\\
    \frac{1}{1+M e^{-\epsilon/T_v}} & \quad x_B = 0
    \end{cases},
\end{equation}
that as noted before and in contrast to the energy switching case, are actually independent of $x_A$. Let us consider now that $T_1 \ll \epsilon \ll T_2$ and that $\lambda_\pm(0;1) \gg \lambda_\pm(0;2)$ and 
$\lambda_\pm(1;1) \ll \lambda_\pm(1;2)$. That means that when the input is $x_A=0$, the output system is more strongly coupled to the cold reservoir at temperature $T_1$, while for input $X_A=1$ the output is more strongly coupled to the hot reservoir at $T_2$. As a consequence, if the frequency of input changes is sufficiently low, for $x_A=0$ the ground state $x_B=0$ will be occupied with higher probability, and for $x_A=1$ the system will be most probably excited. In this way, communication is possible without requiring energy modulation.

The stationary input distribution always satisfies $\pi_{x_A=0}/\pi_{x_A=1} = f_-/f_+$. Note that in this case the channel capacity is not necessarily maximized by a uniform input distribution. Thus, for a fixed average input transition rate $\langle f \rangle \equiv f_+ f_-/(f_- + f_+)$, one needs to find the ratio $f_-/f_+$ that maximizes the channel capacity given all the other parameters. This is done numerically in what follows. Also, we parametrize the transition rates with the constants $\alpha$, $\beta(0)$ and $\beta(1)$ by considering $\lambda_-(0;1) = \lambda_-(1;1) = \alpha \langle f \rangle$ and $\lambda_-(x_A;2) = \beta(x_A) \lambda_-(x_A;1)$. The rest of the rates are determined by \cref{eq:ldb1}. 

\begin{figure}
    \centering
    \includegraphics[scale=.55]{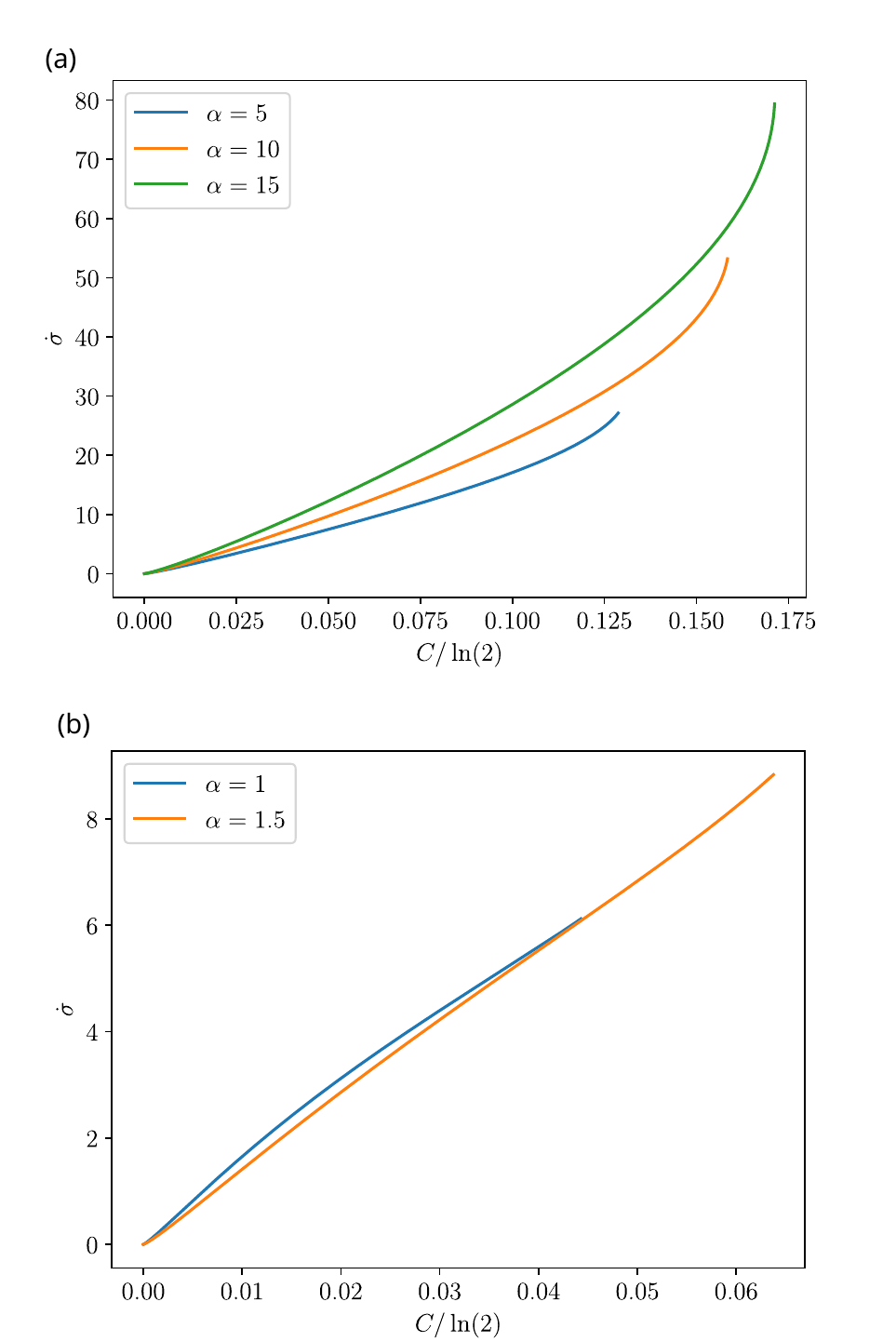}
    \caption{Entropy production rate versus channel capacity for the minimal model of reservoir switching. The parameters are $T_1 /\epsilon = 0.1$, $\beta(1)=\beta(0)^{-1} = 0.1$, and $M=4$. The temperature $T_2$ was varied in the range $(10^{-1}, 10^2)$ to obtain each of the curves.}
    \label{fig:res_switch_ep_vs_cc}
\end{figure}

In Figure \ref{fig:res_switch_ep_vs_cc} we show the entropy production rate versus the channel capacity for different values of $alpha$. All the other parameters are fixed as indicated in the caption with the exception of $T_2$, that was swept to obtain each curve. We see that as we increase $\alpha$, which corresponds to making the output system faster compared to the changing input, it is possible to achieve higher channel capacities at the expense of a higher entropy production rate. For the parameters in Fig. \ref{fig:res_switch_ep_vs_cc}-(a) the entropy production rate is a convex and monotonic function of the channel capacity. 
However, the convexity is lost for lower values of $\alpha$, as shown in Fig. \ref{fig:res_switch_ep_vs_cc}-(b).

\section*{Thermodynamics informs when and how to inverse multiplex}

We have found that in most cases, for both types of communication analyzed above, the EP is a convex function of the channel capacity. 
So by the theory of convex optimization, a thermodynamic benefit to inverse multiplexing must arise. 
In what follows, we analyze when one can reduce thermodynamic cost by splitting information streams across multiple channels. 
We additionally analyze how to distribute information transmission rates across multiple channels in order to reach the minimum achievable thermodynamic cost. 
In doing so, we derive the Pareto front that represents the set of ``optimal'' tuples $(C, \lr{\dot{\sigma}})$ one can achieve for any given fixed number of channels.

\subsection*{$M$-channel capacity}

Consider concurrently operating $M$ channels with capacities $C_1, C_2, \ldots, C_M$; inputs $A_1, \ldots, A_M$; and outputs $B_1, \ldots, B_M$. 
We set the inputs simultaneously at the beginning of each channel use.
We know that the mutual information between the inputs and outputs of $M$ channels is upper bounded by a value that can only be achieved if the inputs are independent~\cite{cover1999infotheory} (Materials and Methods). 
So the capacity of this multi-channel setup equals the sum of the capacities of each channel. 
\eq{
    C &= \max_{p_{X_{A_1},\ldots,X_{A_M}}} I(X_{A_1},\ldots,X_{A_M}; X_{B_1},\ldots,X_{B_M}) \\
    &= \sum_{i=1}^M \max_{p_{X_{A_1},\ldots,X_{A_M}}} I(X_{A_i}; X_{B_i}) 
    % + \max_{p_{X_{A_1}, X_{A_2}}} I(X_{A_2}; X_{b_2}) 
    \\
    &= \sum_{i=1}^M \max_{p_{X_{A_i}}} I(X_{A_i}; X_{B_i}) = \sum_{i=1}^M C_i.
    % + \max_{p_{X_{A_2}}} I(X_{A_2}; X_{b_2}) 
    % \\
    % &= \sum_{i=1}^M C_i
    % \label{eq:s40}
}
The only joint input distribution that achieves this maximum channel capacity is $\pi_{X_{A_1},\ldots,X_{A_M}} = \prod_{i=1}^M \pi_{X_{A_i}}$. 
% Therefore, one achieves the $M$-channel capacity when the joint distributions of the $M$ inputs equals the product distribution of the marginal input distributions that achieve the capacity of each channel. 
So by the channel-coding theorem, optimal codebooks make it appear as if the 
inputs were generated in each channel from its capacity-maximizing input distribution, independently from one another. 
Additionally, since the inputs to all of the channels are statistically independent, there is no unavoidable mismatch cost, as arises for example in the parallel bit erasure of statistically coupled bits~\cite{ouldridge2022thermodynamics}.

\subsection*{$M$-channel EP rate and thermodynamic benefits of inverse multiplexing} 
Since we model the multi-channel setup as if it is running at its capacity, each of the constituent channels is an independent CTMC, governed by its own master equation. 
Therefore, the total EP rate of the combined system is simply the sum of the EP rates of each channel~\cite{wolpert_min_ep_2020}:

\eq{\label{eq:s41}
    \lr{\dot{\sigma}} &= \sum_{i=1}^M \lr{\dot{\sigma}_i} = \sum_{i=1}^M g_i(C_i),
}

\noindent
where $g_i(C_i)$ expresses channel $i$'s EP rate $\lr{\dot{\sigma}_i}$ as a convex function $g_i$ of its channel capacity, $C_i$. 
% \farita{Allocating among N bins - water filling algorithm - convex optimization problem to determine the capacity of each channel so as to minimize the EP rate. }
We ca use the water-filling algorithm to minimize the total EP rate of a set of $M$ channels subject to a desired total channel capacity $C_d = \sum_{i=1}^M C_i$, via the Lagrangian \footnote{This is a convex optimization problem that loosely speaking is the inverse of the one discussed in Section 9.4 of~\cite{cover1999infotheory}, which maximizes total channel capacity subject to a ``power constraint''.}:

\eq{
    \mathcal{L} = \sum_{i=1}^M g(C_i) + \lambda \left( C_d - \sum_{i=1}^M C_i \right)
}

\noindent
Differentiating with respect to $C_j$,

\eq{
    \frac{\partial \mathcal{L}}{ \partial C_j} = g_i'(C_j) - \lambda = 0
}

\noindent
which means

\eq{
    C_j = (g_i')^{-1}(\lambda)
}

\noindent
where $\lambda$ is chosen to satisfy 

\eq{\label{lagrangian_solve}
    \sum_{i=1}^M C_i = \sum_{i=1}^M (g_i')^{-1}(\lambda) = C_d
} 

\noindent
In particular, this suggests that if $g_i = g \, \forall \, i$, then it is optimal to use $M$ channels with identical capacity $\frac{C_d}{M}$. 
%\farita{Double check this intuition / interpretation. }
This finding leads to important insights regarding how to choose between different inverse multiplexing setups. 
For example, a setup of two channels with capacities $C_A$ and $C_B$ is identical in an information-theoretic sense to a setup with two identical channels each with capacity $\frac{1}{2}(C_A + C_B)$. 
However from a thermodynamic viewpoint, it would be better to use the two identical channels because that would minimize EP rate due to the convexity of the EP rate with respect to the channel capacity. 
If we know the functional form of $g(C)$, we can also use the method outlined in~\cite{balasubramanian2015heterogeneity} to calculate the energetically-optimal number of independent, identical channels to use in order to achieve a desired total information rate.  
More generally, for any set of functions $\{g_i\}$, we can find the $\lambda$ that satisfies \cref{lagrangian_solve} (in addition to the positivity of the channel capacity), and plug it into the Lagrangian. 
In this way we can obtain the optimal distribution $\{C_i\}$ of information transmission rates across $M$ channels that are not necessarily identical.

\subsection*{Pareto-optimal fronts for inverse multiplexing}

\begin{figure}[htbp]
    \includegraphics[width=0.49\textwidth]{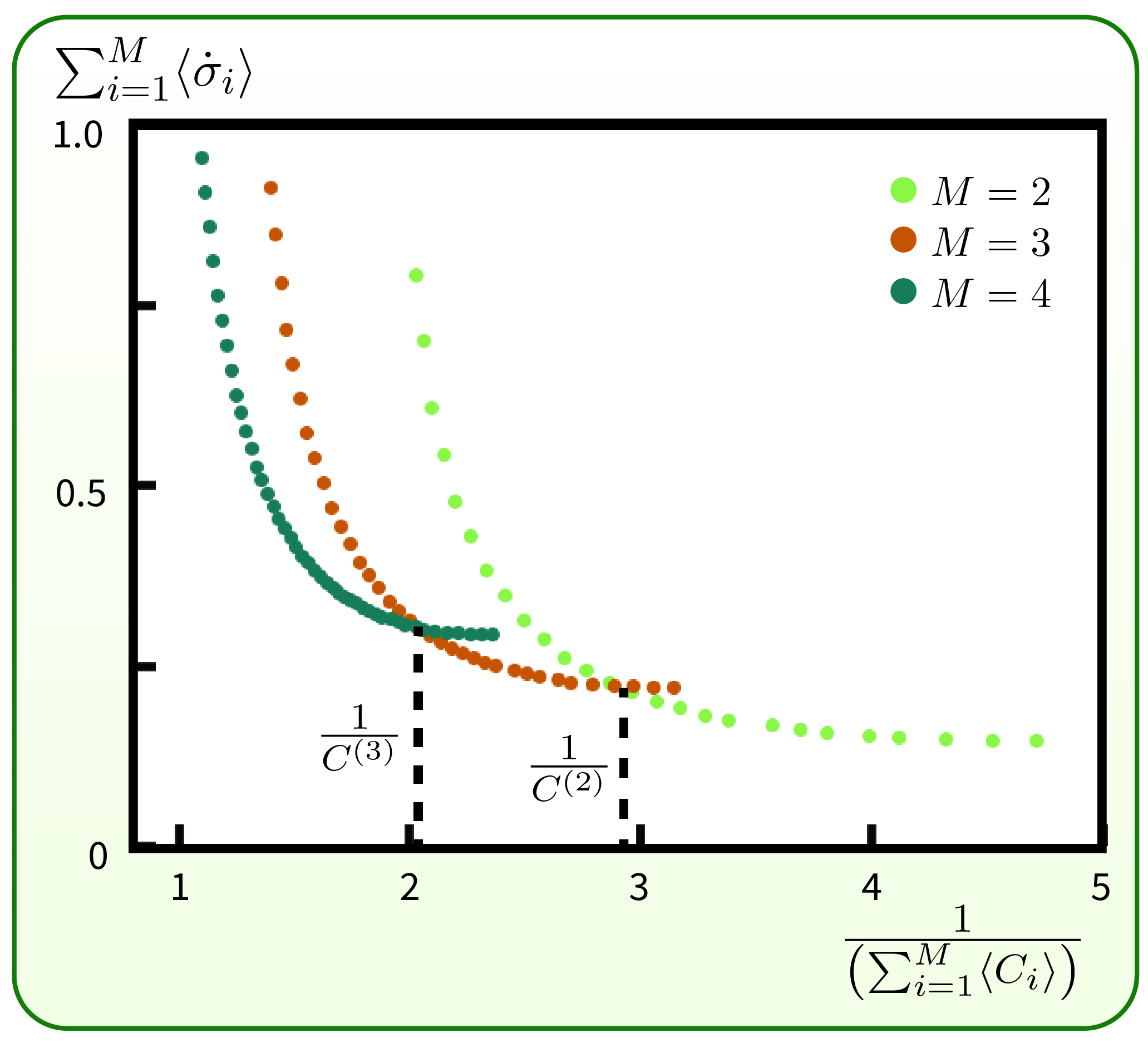}
    \caption{
    \footnotesize
    For a collection of $M$ channels implementing energy switching with capacities $\{C_1, \ldots, C_M \}$, these Pareto fronts are collections of points $(C, \lr{\dot{\sigma}})$ that each minimizes $\lr{\dot{\sigma}} = \sum_{i=1}^M \lr{\dot{\sigma}_i}$ and simultaneously maximizes $C = \sum_{i=1}^M C_i$. 
    All other possible (but non-optimal) combinations $\{C_1, \ldots, C_M \}$ lie to the top right of each front. 
    Above each threshold $C^{(m)}$ splitting information transmission across $m+1$ channels instead of $m$ channels reduces the total EP. 
    In these plots, $f_s = 0.1$, $L=4$, and $\epsilon = 2$. 
    }\label{fig:pareto}
\end{figure}

These results suggest that for any number $M$ of channels, there exists a Pareto front of points $(C, \lr{\dot{\sigma}})$ that each minimizes $\lr{\dot{\sigma}} = \sum_{i=1}^M \lr{\dot{\sigma}_i}$ and simultaneously maximizes $C = \sum_{i=1}^M C_i$. 
We plot examples of these Pareto fronts in \cref{fig:pareto}. 
The plots reveal that higher channel capacities benefit from splitting information rates across more channels. 
More specifically, there exist thresholds, $C^{(m)}$, above which it reduces the total EP to split information transmission across $m+1$ channels instead of $m$ channels. 

%TC:ignore	
%TC:endignore

\section*{Discussion}

% We have shown formally that the EP is a convex and, in many cases, non-monotonic function of the channel capacity. 
% that can be either (i) monotonically increasing and convex; or (ii) quasiconvex with a single global minimum. 
A convex relationship between thermodynamic cost and channel capacity, as well as the energetic benefits of inverse multiplexing that follow, has previously been demonstrated in communication channel models with Gaussian noise~\cite{cover1999infotheory}. 
However, for those models, the ``power'' does not correspond to actual dissipated energy per unit time. 
This relationship has also been justified semi-formally for biological systems~\cite{balasubramanian2001metabolically, balasubramanian2015heterogeneity, balasubramanian2021brainpower}. 
Due to their assumptions, both of these kinds of studies derive that thermodynamic cost increases monotonically with channel capacity. 
To the best of our knowledge, ours is the first study to use physics to demonstrate that, in many cases, the thermodynamic cost is actually not necessarily a monotonic nor convex function of the channel capacity. However, we find that those two properties are recovered for sufficiently large channel capacity, so the thermodynamic benefits of splitting an information stream among multiple channels still arises. 
% Therefore despite previous arguments, it is convexity, not monotonicity, that gives rise to the energetic benefits of inverse multiplexing.  
This paper also presents the first analysis of inverse multiplexing treated with the rigor provided by stochastic thermodynamics. 
% To steer future work in this fundamental area of the thermodynamics of computation, we will now discuss other formulations, how to relax some assumptions, and possibilities for next investigations. 
Our results help illuminate observations in biological communication, and may provide heuristics for an engineer designing communication systems. 
% Our studies represent the first steps in a research effort to rigorously combine nonequilibrium thermodynamics with Shannon information theory. 

There are several  stochastic thermodynamics formulations one ca use to continue to investigate the thermodynamic costs of communication. 
In the main text we use the one in which the dynamics of the system of interest (SOI) is Markovian, due to its coupling with infinite reservoirs that are at thermal equilibrium. 
% Since the reservoirs are infinitely large, they relax to equilibrium much faster than the system does. 
Another formulation, called the ``inclusive Hamiltonian", or ``strong coupling'' model, treats the case where the reservoirs are finite. 
There, the joint system comprising the SOI and the reservoirs evolves with unitary Hamiltonian dynamics \cite{esposito2010entropy, jarzynski2000hamiltonian, ptaszynski2019entropy, timpanaro2019thermodynamic}. 
So the dynamics of the SOI is non-Markovian in general. 
% This formulation equates to that of open quantum thermodynamics if one replaces the marginalizations over the states of the reservoir with partial traces of an environment~\cite{manzano, parrondo, horowitz}. 
There has been a very preliminary analysis of thermodynamically-motivated rate distortion functions for communication channels using this formulation~\cite{kardes2022inclusive}. 

In the main text we have also assumed that the output has ``no back-action'' onto the input, so they each obey LDB separately. 
This non-reciprocity approximates most real-world observable dynamics well~\cite{verley2014nobackaction}. 
However, strictly speaking, many systems violate this approximation when modeled at finer scales due to micro-reversibility and the consequent property that the \textit{entire system} obeys LDB~\cite{wolpert2023combining}. 
% While they are less efficient than channels with no back-action, systems that obey LDB explicitly and exactly can implement communication in principle. 
% The mathematics of the simplest such systems is analytically tractable. 
% We present one such example in the Supplementary Information.
% Ultimately, of course, the precise physical details of the communication channel under consideration, as well as the degree of granularity desired in the model, will determine which stochastic thermodynamics model one should use.

For the purposes of these investigations, we have not considered the thermodynamic cost of the input encoding function and the output decoding function. 
However since different coding strategies (``codebooks'') would likely incur different thermodynamic costs, we suggest that future work should investigate their stochastic thermodynamics. 
Such analyses would naturally extend to investigations of the thermodynamic costs of different error-correcting codes. 

$ $

\noindent \textbf{\large{Acknowledgements}}

\noindent This work was supported by the MIT Media Lab Consortium, Santa Fe Institute, and US NSF EAGER Grant CCF-2221345.

\bibliography{refs.bib}

\appendix

\section{Materials and Methods}

\subsection*{Proof of the form of EP rate versus channel capacity in the energy switching case in the non-equilibrium steady state}

This proof is for the case when $N=2$, $L=2$, and $f_s = 0$. 
Zero signaling rate corresponds to communication protocols in which the input is set once, at time $t=0$, by sampling from $\pi_{X_A}$ remains fixed for the rest of the trajectory, during which only the output can change state. 
While holding $T_2$ fixed and varying $T_1$, we seek to derive the properties of 
\eq{
    \frac{d \lr{\dot{\sigma}}}{d C} = \frac{d \lr{\dot{\sigma}} / dT_1 |_{T_2}}{dC/dT_1|_{T_2}}
}
First, we note that $C$ is always non-negative due to the non-negativity of the mutual information. Its derivative with respect to the temperature of one reservoir
\eq{
    \frac{\partial C}{\partial T_1} &= \frac{\partial p_m}{\partial T_1}(1 + \ln p_m) + \frac{\partial p_e}{\partial T_1}(1 + \ln p_e) \\
    &= \frac{\partial p_m}{\partial T_1} \ln \frac{p_m}{p_e} \\
    &= \frac{1}{r_1 + r_2} \frac{\partial \alpha_1}{\partial T_1} \ln \frac{\alpha}{\zeta} \\
    &= -\frac{\epsilon \alpha_1 \zeta_1}{T_1^2 r_1} \ln \frac{\alpha}{\zeta} \\
    &\leq 0 \; \forall \; T_1 \in (0, \infty)
}
is always non-positive. So the channel capacity is a non-increasing function of the temperature of (any) reservoir. 

We then find that the EP rate can be factorized (assuming the $r_v$ are fixed, positive reals):
\eq{
    \lr{\dot{\sigma}} &= \left( \frac{1}{r_1} + \frac{1}{r_2} \right)^{-1} \left( e^{\frac{\epsilon}{T_1}} - e^{\frac{\epsilon}{T_2}} \right) \left( \frac{\epsilon}{T_1} - \frac{\epsilon}{T_2} \right) \ln \left( \frac{r_1 e^{\frac{\epsilon}{T_1}} + r_2 e^{\frac{\epsilon}{T_2}}}{r_1 + r_2} \right) 
    % \\
    % &:= cf(T_1, T_2)g(T_1, T_2)h(T_1, T_2)
}
Since $\left( e^{\frac{\epsilon}{T_1}} - e^{\frac{\epsilon}{T_2}} \right) \left( \frac{\epsilon}{T_1} - \frac{\epsilon}{T_2} \right) \geq 0$ and the argument of the logarithm is a function that is always $\geq 1$, the EP rate is always non-negative. 
(This is a fact of stochastic thermodynamics as well --- that the ensemble-average EP rate is non-negative.)
Furthermore, it is clear that the EP rate has a singular root that is achieved when $T_1 = T_2$, which gives an EP rate of zero. This singular root must be a minimum since the EP rate is everywhere non-negative.
The EP rate has no other roots, so by Rolle's theorem, 
\eq{
    \frac{d \lr{\dot{\sigma}}}{d T_1}
    \begin{cases}
        < 0 & T_1 < T_2 \\ 
        > 0 & T_1 > T_2
    \end{cases}
}
Therefore,
\eq{ \label{dEPdC}
    \frac{d \lr{\dot{\sigma}}}{d C}
    \begin{cases}
        > 0 & T_1 < T_2 \\ 
        < 0 & T_1 > T_2
    \end{cases}
}
which is exactly what we observe in~\cref{fig:results}. 

We also note the limits
\eq{
    \lim_{T_1 \to 0} C &= \ln 2 - S \left\{ \eta_1 + \eta_2 \left( \frac{e^{\frac{\epsilon}{T_2}}}{e^{\frac{\epsilon}{T_2}} + 1} \right), \eta_2 \left( \frac{1}{e^{\frac{\epsilon}{T_2}} + 1} \right) \right\} \\
    &:= C_{\text{max}}(T_2) \geq 0 \\
    \lim_{T_1 \to \infty} C &= \ln 2 - S \left\{ \frac{\eta_1}{2} + \eta_2 \left( \frac{e^{\frac{\epsilon}{T_2}}}{e^{\frac{\epsilon}{T_2}} + 1} \right), \frac{\eta_1}{2}+ \eta_2 \left( \frac{1}{e^{\frac{\epsilon}{T_2}} + 1} \right) \right\}  \\
    &:= C_{\text{min}}(T_2) \geq 0
}
where $\eta_1 = \frac{r_1}{r_1+r_2}$ and $\eta_2 = \frac{r_2}{r_1+r_2}$. 
So, the channel capacity has minimum and maximum values. 
We also find that, at the minimum, the channel capacity equals
\eq{
    C(T_1 = T_2 = T) = \ln 2 - S \left(\left\{ \frac{e^{\frac{\epsilon}{T}}}{e^{\frac{\epsilon}{T}} + 1} , \frac{1}{e^{\frac{\epsilon}{T}} + 1} \right\} \right)
}
which is strictly positive unless $T = \infty$.
Furthermore, we write the limits on the entropy production rate
\eq{
    \lim_{T_1 \to 0} \lr{\dot{\sigma}} &= \infty \\
    \lim_{T_1 \to \infty} \lr{\dot{\sigma}} &= \left( \frac{1}{r_1} + \frac{1}{r_2} \right)^{-1} \frac{\epsilon}{T_2}\left( e^{\frac{\epsilon}{T_2}} - 1 \right) \ln \left( \frac{r_1 + r_2 e^{\frac{\epsilon}{T_2}}}{r_1 + r_2} \right) \\
    &:= \lr{\dot{\sigma}}^\dagger(T_2) \geq 0
}

Putting \cref{dEPdC} together with these limit analyses results in the fact that the EP rate as a function of the channel capacity is convex with a single global minimum that occurs at a positive value of the channel capacity. 
(The channel capacity is zero when $T_1 = T_2 = \infty$.) 
Additionally, the EP rate diverges as the channel capacity approaches its maximum value $C_{\text{max}}(T_2)$, as depicted in \cref{fig:results}. 
So for this energy-switching case, the EP is a convex, non-monotonic function of the channel capacity.

\subsection*{Mutual information is maximized when channels are independent}

For any number $M \geq 2$ of channels~\cite{cover1999infotheory},
\eq{
	I&(X_{A_1}, \ldots, X_{A_M}; X_{B_1}, \ldots, X_{B_M}) \\
	&= H(X_{B_1}, \ldots, X_{B_M}) - H(X_{B_1}, \ldots, X_{B_M} | X_{A_1}, \ldots, X_{A_M}) \\
	&= H(X_{B_1}, \ldots, X_{B_M}) - \sum_{i=1}^M H(X_{B_i}| X_{A_1}, \ldots, X_{A_M}) \\
	&= H(X_{B_1}, \ldots, X_{B_M}) - \sum_{i=1}^M H(X_{B_i}| X_{A_i}) \\
	&\leq \sum_{i=1}^M H(X_{B_i}) - \sum_{i=1}^M H(X_{B_i}| X_{A_i}) \label{eq:dualMI} \\
	&\qquad = \sum_{i=1}^M I(X_{A_i}; X_{B_i})
}
\cref{eq:dualMI} achieves equality when the $M$ channels are all independent.

\end{document}